\documentclass[preprint,aps,showpacs,epsf,tighten]{revtex4}	
\usepackage{graphicx}
\def\lsim{\buildrel < \over {_{\sim}}}
\def\gsim{\buildrel > \over {_{\sim}}}
\begin{document}
%
%
\vspace*{-0.5in}
          \hfill \begin{minipage}[t]{2in}{
	  {FERMILAB-Pub-05-021-T}\\
	  hep-ph/0502161\\
	  May 2005\\[0.2in]}
	  \end{minipage}

\title{	Bounds on the neutrino mixing angles and CP phase\\
          for an $SO(10)$ model with lopsided mass matrices\\[0.3in]}
\author{Carl H. Albright}
\email[Electronic address: ]{albright@fnal.gov}
\affiliation{Department of Physics, Northern Illinois University, 
	DeKalb, IL 60115}
\affiliation{Fermi National Accelerator Laboratory, P.O. Box 500,
	Batavia, IL 60510}
\begin{abstract}

The bounds on the neutrino mixing angles and CP Dirac phase for an $SO(10)$ 
model with lopsided mass matrices, arising from the presence of ${\bf 16}_H$ 
and $\overline{\bf 16}_H$ Higgs representations, are studied by variation of 
the one real and three unknown complex input parameters for the right-handed 
Majorana neutrino mass matrix.  The scatter plots obtained favor nearly 
maximal atmospheric neutrino mixing,  while the reactor neutrino mixing 
lies in the range $10^{-5} \lsim \sin^2 \theta_{13} \lsim 1 \times 10^{-2}$ 
with values greater than $10^{-3}$ most densely populated.  A rather compelling
scenario within the model follows, if we restrict the three 
unknown complex parameters to their imaginary axes and set two of them equal.
We then find the scatter plots are reduced to narrow bands, as the mixing 
angles and CP phase become highly correlated and predictive.  The bounds on 
the mixing angles and phase then become $0.45 \lsim \sin^2 \theta_{23} 
\lsim 0.55$,\ $0.38 \lsim \tan^2 \theta_{12} \lsim 0.50$,\ $0.002 \lsim 
\sin^2 \theta_{13} \lsim 0.003$,\ and $60^\circ \lsim \pm \delta_{CP} 
\lsim 85^\circ$.  Moreover, successful leptogenesis and subsequent 
baryogenesis are also obtained, with $\eta_B$ increasing from  $(2.7\ {\rm to}
\ 6.3) \times 10^{-10}$ as $\sin^2 \theta_{23}$ increases from 0.45 to 0.55.

\end{abstract}

\maketitle
\thispagestyle{empty}
\newpage
%

\section{INTRODUCTION}

Since the discovery of atmospheric neutrino oscillations \cite{atm}, more 
recently that of solar neutrino oscillations \cite{sol}, and still more 
recently confirmation of these two types of neutrino oscillations with 
accelerator- and reactor-produced neutrinos \cite{accel,reac}, many models 
have been proposed in the literature to explain the mass and 
mixing parameters associated with these oscillations \cite{models}.  
(We have cited just the most recent references for those experiments and 
several review articles for the models proposed.)  Some models restrict 
their scope to the lepton sector, while others such as grand unified models 
are more ambitious and attempt to explain the masses and mixings in both 
the lepton and quark sectors.  The textures for the mass 
matrices obtained in the models may be simply postulated at the outset, 
derived from the observed mixing matrix with a diagonal charged lepton 
mass matrix, obtained from a certain unification group in conjunction with
a well-defined family symmetry, or rely solely on the Clebsch-Gordon 
coefficients of that group.  
 
Until now, all viable models for the lepton sector have only had to explain 
the apparent near maximal $\theta_{23}$ mixing angle for the atmospheric 
neutrinos and the less than maximal $\theta_{12}$ large mixing angle (LMA) 
solution for the solar neutrinos, while satisfying the two observed mass 
squared differences, $\Delta m^2_{32}$ and $\Delta m^2_{21}$, and 
the upper bound for the still unobserved $\theta_{13}$ mixing angle. On the 
other hand, the neutrino mass hierarchy (normal or inverted), the Majorana 
vs Dirac nature of the neutrinos, the appropriate values for $\theta_{13}$ 
and the CP-violating phases (one Dirac and two Majorana) remain 
undetermined experimentally.  

With the next more precise round of neutrino oscillation experiments which 
will restrict some of the unknowns, many 
of the more quantitative models proposed are not expected to survive the more 
stringent tests of their predictions. In particular, determination of the 
neutrino mass hierarchy will eliminate models based on an approximately 
conserved $L_e - L_\mu - L_\tau$ lepton number \cite{consL} if the hierarchy 
is observed to be normal, while the conventional type I seesaw models will 
be strongly disfavored if the hierarchy is observed to be inverted 
\cite{typeI}.  Moreover, small as opposed to very small predictions for 
$\theta_{13}$ pose a critical test for many models.  It is thus of utmost 
importance to sharpen the predictions for each model, so that the list of 
presently viable models can be winnowed down as much as possible when the 
new data is obtained. 

One particularly interesting class of models is that based on the $SO(10)$
grand unification group.  There are two kinds of minimal models in this class:
those based on Higgs representations with dimensions ${\bf 10},\ {\bf 126},
\ {\overline{\bf 126}}$, and possibly also ${\bf 120}$ and/or ${\bf 210}$
\cite{minimal}; and those based on ${\bf 10},\ {\bf 16},\ 
{\overline{\bf 16}}$, and ${\bf 45}$ representations \cite{lopsided}.  The 
former type generally has symmetric and/or antisymmetric matrix elements, 
while the latter type generally involve lopsided matrices for the 
down quarks and charged leptons.  The matrix textures are derived either 
from the Clebsch-Gordon coefficients required or from some Abelian or 
non-Abelian family symmetry.  These two types of $SO(10)$ models tend to 
predict $\theta_{13}$ angles either near the present upper CHOOZ bound of
$12^\circ$, or in the $1^\circ - 3^\circ$ range, respectively.

The purpose of this paper is to sharpen the allowed values for the mixing 
parameters of the lopsided $SO(10)$ model proposed some time ago by Babu,
Barr, and the author \cite{abb}.  Upon further refinement by the latter
two authors \cite{abLMA}, the model has become very predictive with all four 
quark mixing parameters and nine quark and charged lepton masses determined by
just eight of the twelve model input parameters.  The four remaining input
parameters characterize the right-handed Majorana mass matrix with three of  
them complex in general.  In most of the previous papers based on this model,
the authors took the latter parameters to be real and adjusted them to give 
good fits to the then known neutrino mass and mixing data.  Here we allow all 
three to be complex and obtain scatter plots of the predicted neutrino mixing 
parameters.  In so doing, we obtain some bounds on $\theta_{13}$ and 
$\theta_{23}$ which, if violated, will rule out the model which has been 
highly successful todate.  By introducing two very small additional parameters
in the Dirac neutrino mass matrix, we find that the observed baryon asymmetry
can be obtained through resonant leptogenesis in the model.  In order to 
obtain the latter predictions, the bounds obtained on the neutrino mixing
parameters and CP phase are even more tightly constrained.  With maximal 
CP violation in the Majorana mass matrix and two of the three parameters 
equal, the mixing angles and phase become highly correlated with each other
and the baryon asymmetry generated in the model.

In Sec. II we briefly review some of the model details and present the 
results of the parameter searches in Sec. III.  A summary of the constraints
found in the model is presented in Sec. IV, where we also briefly comment on 
the predictions of the model found by Jankowski and Marbury \cite{jm} for the 
$\mu \rightarrow e + \gamma$ lepton flavor-changing branching ratio. 
Their prediction is also crucial for the viability of the model and has a 
chance of being tested by the MEG collaboration \cite{MEG} even before the 
values of the mixing angles are further restricted by new reactor and/or long 
baseline experiments.

\section{$SO(10)$ MODEL and ITS ARBITRARY PARAMETERS}

The $SO(10)$ grand unified model as initially proposed by Babu, Barr and the 
author \cite{abb}, and later updated by Barr and the author to apply to the 
large mixing angle (LMA) neutrino oscillation solution has already been 
described in great detail in the literature \cite{abLMA}.  Here we just 
summarize the salient features and refer the reader to those references for 
more information.

The model is based on the minimum set of Higgs fields which solves the 
doublet-triplet splitting problem as proposed by Barr and Raby \cite{br}.  
This requires just one ${\bf 45}_H$ whose vacuum expectation value (VEV) 
points in the $B-L$ direction, 
two pairs of ${\bf 16}_H,\ {\bf \overline{16}}_H$'s which stabilize the 
solution, along with several Higgs fields in the ${\bf 10}_H$ representations 
and Higgs singlets.  The Higgs superpotential exhibits the $U(1) \times Z_2 
\times Z_2$ symmetry which is used for the flavor symmetry of the grand
unified model.  
The combination of VEVs, $\langle {\bf 45}_H\rangle_{B-L},\ 
\langle 1({\bf 16}_H)\rangle$ and $\ \langle 1({\bf \overline{16}_H})\rangle$ 
break $SO(10)$ to the Standard Model, where the latter two VEVs point in the 
$SU(5)$ singlet direction.  The electroweak VEVs arise from doublets in the 
$5$ and $\bar{5}$ representations of $SU(5)$ in the combinations
\begin{equation}
\begin{array}{rcl} 
        \langle H_u \rangle &=& \langle 5({\bf 10}_H)\rangle,\\[0.1in]
        \langle H_d \rangle &=& \langle \overline{5} ({\bf 10}_H)\rangle
          \cos \gamma + \langle \overline{5}({\bf 16}'_H)\rangle \sin \gamma,\\
\end{array}
\label{eq:Higgsmixings}
\end{equation}
while the mixing orthogonal to $H_d$ gets massive at the grand unified theory 
(GUT) scale.  Matter superfields appear 
in three chiral ${\bf 16}$'s, two pairs of vectorlike ${\bf 16}$ and 
$\overline{\bf 16}$, two ${\bf 10}$'s and several ${\bf 1}$'s.
The mass matrices follow from Froggatt-Nielsen diagrams \cite{fn} in which the 
superheavy fields are integrated out, with the flavor symmetry allowing only 
certain particular contributions to their mass matrix elements. 
The Dirac mass matrices for the up quarks, down quarks, neutrinos and charged 
leptons are found to be
\begin{equation}
\begin{array}{ll}
M_U = \left(\matrix{ \eta & \delta_N & \delta'_N \cr
  \delta_N & 0 & - \epsilon/3 \cr \delta'_N & \epsilon/3 & 1\cr} \right)m_U,\
  & M_D = \left(\matrix{ 0 & \delta & \delta' e^{i\phi}\cr
  \delta & 0 & - \epsilon/3  \cr
  \delta' e^{i \phi} & \sigma + \epsilon/3 & 1\cr} \right)m_D, \\[0.5in]
M_N = \left(\matrix{ \eta & \delta_N & \delta'_N \cr \delta_N & 0 & \epsilon 
	\cr  \delta'_N & - \epsilon & 1\cr} \right)m_U,\
  & M_L = \left(\matrix{ 0 & \delta & \delta' e^{i \phi} \cr
  \delta & 0 & \sigma + \epsilon \cr \delta' e^{i\phi} &
  - \epsilon & 1\cr} \right)m_D.\\
\end{array}
\label{eq:Dmatrices}
\end{equation}
in the convention where the left-handed fields label the rows and the 
left-handed conjugate fields label the columns.  

The entries in the first row and first column are the most uncertain, 
especially for the up quark and Dirac neutrino mass matrices, for they arise
from higher order terms involving several integrations out of the massive
neutrino lines. As originally proposed, the parameters $\delta_N$ and 
$\delta'_N$ were absent.  In a later application of the model to leptogenesis 
\cite{lepto}, we have observed that the baryon asymmetry lies closer to the 
preferred value of $\eta_B \simeq 6.2 \times 10^{-10}$ \cite{etaB}, if we 
allow $\delta_N$ and $\delta'_N$ to take on nonzero values in the Dirac 
neutrino matrix $M_N$, which are small enough in $M_U$, however, so as not to 
upset the good results in the quark sector.  We shall consider both  
possibilities for these two parameters in the searches to be presented.

The lopsided nature of the down quark and charged lepton mass matrices arises 
from the dominance of the $\sigma$ ${\bf 16}'_H$ contribution to the 
electroweak symmetry breaking over that of the ${\bf 10}_H$.  The pronounced 
lopsidedness with $\tan \beta \simeq 5$ readily explains the small $V_{cb}$ 
quark mixing and near maximal $U_{\mu 3}$ atmospheric neutrino mixing for 
any reasonable right-handed Majorana neutrino mass matrix, $M_R$.  
All nine quark and charged lepton masses, plus the three CKM angles and CP
phase, are well-fitted with the original eight input parameters defined at 
the GUT scale to fit the low scale observables after evolution 
downward from $\Lambda_{GUT} \simeq 2 \times 10^{16}$ GeV \cite{newinput}:
\begin{equation}
\begin{array}{rlrl}
        m_U&\simeq 113\ {\rm GeV},&\qquad m_D&\simeq 1\ {\rm GeV},\\
        \sigma&= 1.83,&\qquad \epsilon&=0.147,\\
        \delta&= 0.00946,&\qquad \delta'&= 0.00827,\\
        \phi&= 119.4^\circ,&\qquad \eta&= 6 \times 10^{-6},\\
\end{array}
\label{eq:inputparam}
\end{equation}

The effective light neutrino mass matrix, $M_\nu$, is obtained from the 
conventional type~I seesaw mechanism \cite{seesaw} once the right-handed 
Majorana mass matrix, $M_R$, is specified.  While the large atmospheric 
neutrino mixing $\nu_\mu \leftrightarrow \nu_\tau$ arises primarily from 
the structure of the charged lepton mass matrix $M_L$, the structure of 
the right-handed Majorana mass matrix $M_R$ determines
the type of $\nu_e \leftrightarrow \nu_\mu,\ \nu_\tau$ solar neutrino mixing,
so that the solar and atmospheric mixings appear to be essentially decoupled 
in the model.  The LMA solar neutrino solution is obtained with a special 
form of $M_R$ \cite{abLMA}, which can be explained by the structure of the 
Froggatt-Nielsen diagrams:  
\begin{equation}
          M_R = \left(\matrix{c^2 \eta^2 & -b\epsilon\eta & a\eta\cr
                -b\epsilon\eta & \epsilon^2 & -\epsilon\cr
                a\eta & -\epsilon & 1\cr}\right)\Lambda_R,\\
\label{eq:MR}
\end{equation}
where the parameters $\epsilon$ and $\eta$ are those introduced in 
Eq. (\ref{eq:Dmatrices}) for the Dirac sector.  The vanishing $2-3$ 
subdeterminant of $M_R$ arises naturally if the Yukawa couplings are
universal at the GUT scale.  The new parameters $a,\ b$ and $c$ are 
undetermined but expected to be of $\mathcal{O}(1)$, since the first row and 
first column of $M_R$ have been properly scaled by powers of the very 
small $\eta$ parameter.  As such, the right-handed Majorana mass matrix
exhibits a very strong hierarchy.  

Given the right-handed Majorana mass matrix above and $\delta_N = \delta'_N 
= 0$, the seesaw formula results in 
\begin{equation}
\begin{array}{rcl}
	M_\nu &=& - M_N M^{-1}_R M^T_N\\[0.1in] 
		&=& -\left(\matrix{ 0 & 
                        \frac{1}{a-b} \epsilon & 0\cr 
                        \frac{1}{a-b} \epsilon & \frac{b^2-c^2}{(a-b)^2} 
                        \epsilon^2 
                        & \frac{b}{b-a} \epsilon\cr 0 & \frac{b}{b-a} \epsilon 
                        & 1\cr} \right)m^2_U/\Lambda_R.
\end{array}
\label{eq:MnuI}
\end{equation}
Note that the very strong hierarchy in $M_R$ can nearly balance the large
hierarchy in $M_N$ to yield a rather mild mass hierarchy for the 
$M_\nu$ light neutrino mass matrix.  In fact, if $M_R \propto M^T_N M_N$ were 
to hold, which has a texture similar to that in Eq. (\ref{eq:MR}), the seesaw 
formula would yield a light neutrino mass matrix exactly proportional to 
the identity matrix resulting in a three-fold mass degeneracy.  The author 
has argued in Ref. \cite{typeI} that a normal hierarchy solution is much 
more stable than an inverted hierarchy solution for such type I seesaw models,
and our results apply for this kind of solution.  Barr has recently proposed 
more general possibilities which result in such a mild neutrino mass hierarchy 
\cite{barr}.

The parameter $\Lambda_R$ can be chosen real and to a large degree sets the 
scale for the atmospheric neutrino mixing mass squared difference.  In most 
previous studies of the model, aside from that of leptogenesis, we chose the 
remaining three parameters, $a,\ b$ and $c$, to be real.  In order to 
determine the range of predictions for the neutrino mixing angles, we now 
allow all three of them to be complex.

\section{\bf RESULTS OF THE PARAMETER SEARCH}

In order to carry out the search of the parameter space for a stable normal
hierarchy solution, we shall require that the oscillation parameters 
determined by the model lie in the following presently allowed ranges
at the 90\% confidence level \cite{atm}-\cite{reac}, \cite{CHOOZ}:
\begin{equation}
\begin{array}{rlrl}
	\Delta m^2_{32} &= (1.9 - 3.0) \times 10^{-3}\ {\rm eV^2},&
		\quad \sin^2 \theta_{23}&= 0.36 - 0.64,\\[0.1in]
	\Delta m^2_{21} &= (7.4 - 8.5) \times 10^{-5}\ {\rm eV^2},&
                \quad \tan^2 \theta_{12} &= 0.33 - 0.50,\\[0.1in]
	\Delta m^2_{31} &\simeq \Delta m^2_{32},& \quad \sin^2 2\theta_{13} 
		&\leq 0.16.\\
\label{eq:nudata}
\end{array}
\end{equation}
The allowed variation in $\sin^2 \theta_{23}$ corresponds to the present 
bound of $\sin^2 2\theta_{23} \geq 0.92$ \cite{atm}.  We find that the 
following bounds for the magnitudes of the input parameters in $M_R$ cover
the above experimentally allowed ranges:
\begin{equation}
	|a| \leq 2.4, \qquad |b| \leq 3.6, \qquad |c| \leq 3.6, 
\label{eq:bounds}
\end{equation}
where we have set $\Lambda_R = 2.85 \times 10^{14}$ GeV as explained below.  
We then use Monte Carlo techniques to throw points in the complex $a,\ b$ 
and $c$ planes which uniformly cover the disks whose radii are indicated above.

It should be noted that in this paper, we shall neglect any running of the 
neutrino mass and mixing parameters from the GUT scale down to the low scales.
This has become more or less standard procedure in the literature and is 
less susceptible to corrections when $\tan \beta$ is low, the neutrino mass
hierarchy is normal, and the two low-lying neutrino mass eigenstates have 
nearly opposite $CP$-parities \cite{radcorr}.  All three conditions hold in 
the model under consideration.  In fact, we can use the convenient add-on
program package for Mathematica called ``REAP'' provided by Antusch et al. 
\cite{aklrs} to test the assumption that the evolution effects are negligible 
for our case.

For this purpose, what is required in addition to the mass matrices in Eq. 
(\ref{eq:Dmatrices}) is specification of the appropriate Higgs VEVs, 
$v_u \equiv \langle 5({\bf 10}_H) \rangle$ and $v_d \equiv \langle 
\bar{5}({\bf 10}_H \rangle$ in Eq. (\ref{eq:Higgsmixings}), together with the 
two input scaling masses, $m_U$ and $m_D$; from these the corresponding Yukawa 
coupling matrices can be determined.  Since $\langle H_u \rangle$ receives a 
contribution only from $\langle 5({\bf 10}_H) \rangle$, the 33 element of the 
Dirac matrix $M_N$ is given by 
\begin{equation}
	(M_N)_{33} = m_U = (Y_N)_{33}v_u,       
\label{eq:vu}
\end{equation}
where $Y_N$ is the Yukawa coupling matrix for the Dirac neutrino mass matrix 
and $v_u = v\sin \beta$ with $v = 174$ GeV as usual.  On the other hand, 
$\langle H_d \rangle$ has two contributions from $\langle \bar{5}({\bf 10}_H) 
\rangle$ and $\langle \bar{5}({\bf 16}_H) \rangle$, but only the first 
contributes to the 33 element of the charged lepton mass matrix $M_L$; hence
\begin{equation}
	(M_L)_{33} = m_D = (Y_L)_{33}v_d \cos \gamma,
\label{eq:vd}
\end{equation}
where $Y_L$ is the Yukawa coupling matrix for the charged leptons and $\cos 
\gamma$ is the projection of $H_d$ onto the VEV of $\langle 
\bar{5}({\bf 10}_H) \rangle \equiv v_d = v\cos \beta$.  With $\tan \beta = 
v_u/v_d = 5$ as suggested earlier to provide a sufficiently lopsided charged 
lepton mass matrix, and the values for $m_U$ and $m_D$ in Eq. 
(\ref{eq:inputparam}), we find that
\begin{equation}
\begin{array}{rcl}
	Y_N &=& 0.662 (M_N/m_U),\\[0.1in]
	Y'_L &\equiv& Y_L \cos \gamma = 0.0293 (M_L/m_D),
\end{array}
\label{eq:Yukawas}
\end{equation}
where we have defined $Y'_L$ as the effective charged lepton Yukawa coupling 
matrix due to the $\cos \gamma$ suppression.  We assume that the VEVs, 
$v_u$ and $v_d$, are held constant while $Y_N$ and $Y'_L$ evolve from the 
GUT scale to the weak scale according to the renormalization group equations.

By inserting this information into the REAP package of Antusch et al. 
\cite{aklrs}, we can obtain the evolved results for the neutrino mixing 
parameters.  We find for 
various test cases that the evolved mixing angles and phases are changed by 
less than $1\%$ from their GUT scale values, while the light neutrino masses
are reduced by $20\%$.  The latter discrepancy can be corrected if we simply
reduce the right-handed Majorana scaling factor $\Lambda_R$ by $20\%$ from
$2.85 \times 10^{14}$ GeV to $2.25 \times 10^{14}$ GeV, with essentially no
alteration in the mixing angles and phases.  With this understanding of the 
role played by evolution of the Yukawa couplings, we simply ignore the 
evolution and proceed to use the GUT scale parameters given earlier.  
	
Scatter plots are shown in Fig. 1 for the original choice of $\eta = 0.6 
\times 10^{-5}$ and $\delta_N = \delta'_N = 0$. The atmospheric and solar 
neutrino mixing regions are shown in Figs. 1(a) and (b), where 
$\Delta m^2_{32}$ and $\Delta m^2_{21}$ are plotted against 
$\sin^2 2\theta_{atm} = 4|U_{\mu 3}|^2|U_{\tau 3}|^2$ and 
$\sin^2 2\theta_{sol} = 4|U_{e1}|^2|U_{e2}|^2$, respectively, the atmospheric
and solar mixing angles in terms of the neutrino mixing matrix elements.  For 
the atmospheric neutrino mixing, we 
observe from Fig. 1(a) that while $\Lambda_R$ sets the scale for the mass 
squared difference, there is a spread of points for $\Delta m^2_{32}$ in the 
allowed interval $(1.9,3.0) \times 10^{-3}\ {\rm eV^2}$; cf. Eq. (5).  
A value of $\Lambda_R = 2.85 \times 10^{14}$ GeV has been chosen, in order 
to center the allowed region predicted by the model on the experimentally 
allowed region as determined by the Super Kamiokande data.  For their best 
fit point of $\Delta m^2_{32} = 2.4 \times 10^{-3}\ \rm{eV}^2$ and $\sin^2 
2\theta_{23} = 1.00$, the model prefers a nearly maximal mixing angle of 
$\sin^2 2\theta_{23} \gsim 0.98$, in agreement with their finding.  Note that 
the highest concentration of model points also occurs in this region 
surrounding their best fit point.  For the solar neutrino mixing depicted in 
Fig. 1(b), we observe a nearly uniform spread in $\Delta m^2_{21}$ over the 
region allowed in Eq. (\ref{eq:nudata}) by the present data.  On the other 
hand, for the solar mixing angle there is a slightly greater concentration of 
points toward the smaller allowed values for $\sin^2 2\theta_{sol}$. 

Although it has been customary to present the neutrino mixing data in 
$\Delta m^2$ vs. $\sin^2 2\theta$ mixing planes as in Figs. 1(a) and (b), 
some authors \cite{sinang} have suggested that $\sin^2 \theta$ is a more 
sensitive measure of the mixing angle, especially for angles close to a 
maximal mixing of $45^\circ$.  Of particular interest is whether 
$\theta_{23}$ lies below $45^\circ$ or above $45^\circ$, if the mixing is not 
exactly maximal.  With a normal mass hierarchy as suggested by our model, the 
former signifies that the highest neutrino mass eigenstate corresponds more 
to a $\nu_\tau$ flavor composition rather than to a $\nu_\mu$ composition, 
while the opposite is true for the latter possibility.   In Figs. 1(c)-(f) 
we also present results for (c) $\tan^2 \theta_{12}$, (d) $\sin^2 \theta_{13}$,
(e) $\delta_{CP}$, and (f) the baryon asymmetry, $\eta_B$, vs. 
$\sin^2 \theta_{23}$.  We see from these plots that there is just a slight 
preference for values of $\sin^2 \theta_{23}$ greater than 0.5 than for those 
less than 0.5.  Concerning the solar neutrino mixing angle distribution in 
Fig. 1(c), the full range of $0.33 \leq \tan^2 \theta_{12} \leq 0.50$ is 
populated, but values below 0.4 are slightly preferred to those above 0.4 in 
keeping with remarks made about Fig. 1(b). Nearly all the allowed points for 
$\sin^2 \theta_{13}$ in Fig. 1(d) are found to lie below 0.012 and above 
$1 \times 10^{-5}$, with those values above 0.001 most densely populated.  
Interestingly enough, the present best fits for $\sin^2 \theta_{13}$ are 
0.006 and 0.004 from three-neutrino mixing analyses with all oscillation data 
considered \cite{best}.  The model seems to be rather insensitive to the 
leptonic CP-violating phase $\delta_{CP}$ in this more general search of 
the complex values for $a,\ b$ and $c$, for the full range from $-90^\circ$ 
to $90^\circ$ is essentially uniformly populated in Fig. 1(e).  
Finally we note that values for the baryon asymmetry in Fig. 1(f) are 
concentrated in the $10^{-13} - 10^{-12}$ range, which are much too small to 
explain the observed baryon asymmetry in the universe.  

In Fig. 2, the corresponding scatter plots are presented for the choice 
$\eta = 1.1 \times 10^{-5},\ \delta_N = -1.0 \times 10^{-5}$ and 
$\delta'_N = -1.5 \times 10^{-5}$.  This latter choice has been made as a 
result of a protracted search, so as to maximize the calculated baryon 
asymmetry as we shall see, without spoiling the results in the quark sector.  
In the previous study reported in the last paper of reference \cite{lepto}, 
the corresponding values taken were $\eta = 0.6 \times 10^{-5},\ 
\delta_N = -0.65 \times 10^{-5}$ and $\delta'_N = -1.0 \times 10^{-5}$.
But without altering the parameter, $\eta$, we observed a maximum of 
$\eta_B$ which fell short of the presently observed baryon asymmetry.

The atmospheric neutrino mixing plane in Fig. 2(a) exhibits a somewhat 
stronger preference for lower values of $\Delta m^2_{32}$ than in Fig. 1(a),
while the solar mixing angle distributions in Figs. 2(b) and (c) exhibit a 
slightly greater preference for smaller solar mixing angles. From Fig. 2(d)
we see the upper bound on $\sin^2 \theta_{13}$ is predicted to lie lower than 
before in Fig. 1(d) and is now given by $\sin^2 \theta_{13} \lsim 0.007$.  
The distribution is also somewhat more diffuse and less concentrated near the 
upper limit.  Again no preferred values of the CP-violating phase 
$\delta_{CP}$ are observed in Fig. 2(e), but the baryon asymmetry is shifted 
to higher values in the broader range, $3 \times 10^{-13}$ to $5 \times 
10^{-11}$ in Fig. 2(f).

The scatter plots presented in Figs. 1 and 2 were obtained by throwing 
400,000 sets of points for the parameters $a,\ b$ and $c$ in the ranges 
given in Eq. (\ref{eq:bounds}) and demanding that the calculated mixing 
parameters lie in the experimental ranges quoted in Eq. (\ref{eq:nudata}).
But the question arises whether more sparsely scattered points can be 
obtained outside the apparent boundaries.  This is of special interest for 
the baryon asymmetry, where in fact one isolated point in Fig. 2(f) does show 
up with $\eta_B = 2.6 \times 10^{-10}$, well outside the major clustering but
still below the observed value of $6.2 \times 10^{-10}$.  Rather than throw
10 - 100 times as many points, we searched for $M_R$ input parameters
which favor larger values of $\eta_B$.  We found such a region defined
by constraining the parameters $a,\ b$ and $c$ to lie near or along their 
imaginary axes in the complex planes, still bounded by the magnitudes 
given in Eq. (\ref{eq:bounds}).

In Figs. 3(a) and (b), we present the scatter plots in the atmospheric and
solar neutrino mixing planes obtained by throwing 100,000 sets of points 
for positive and negative imaginary values of $a,\ b$ and $c$ bounded by
the magnitudes of Eq. (\ref{eq:bounds}), again for the new values of 
$\eta,\ \delta_N$ and $\delta'_N$.  While the solar mixing plane of
Fig. 3(b) is very similar to that of Fig. 2(b), the atmospheric mixing plane
in Fig. 3(a) is dramatically affected as the spread of points now lie along 
the arc shown.  The upper branch corresponds to values of 
$\sin^2 \theta_{23} < 0.5$ and the lower branch to values of 
$\sin^2 \theta_{23} > 0.5$.  The small secondary upper branch is obtained 
when $|c| < |a|$, while the major upper branch is found when $|c| > |a|$.  
Further insight into the restricted results are obtained from Figs. 3(c)-(f).
Bands develop in all four plots of (c) $\tan^2 \theta_{12}$, 
(d) $\sin^2 \theta_{13}$, (e) $\delta_{CP}$, and (f) $\eta_B$ vs.
$\sin^2 \theta_{23}$.  One of the most striking results is that 
$\sin^2 \theta_{13}$ is now limited to the restricted range (0.0018, 0.0035).
As seen in Fig. 3(f), 29~points have appeared in the $\eta_B > 1.0 \times 
10^{-10}$ range.

In Fig. 4 similar plots for the mixing planes and distributions are shown,
where 200,000 sets of points for imaginary values of $a,\ b$ and $c$ were 
thrown with the restriction $\eta_B \geq 1.0 \times 10^{-10}$ applied.  In 
all 84 points survive this cut.  
We now see that the upper branch in the atmospheric neutrino mixing plane 
of Fig. 4(a) is essentially eliminated, for $\sin^2 \theta_{23} \gsim 0.47$ 
is required for most points to obtain such large values of $\eta_B$.  The 
scatter points in the solar neutrino mixing plane is little affected aside 
from the density of points as seen in Fig. 4(b).  The bands in Figs. 3(c), 
(d) and (e) become extremely narrow in Figs. 4(c), (d), and (e) when the cut 
$\eta_B \geq 1.0 \times 10^{-10}$ is imposed.   A strong correlation arises
between the solar mixing angle in $\tan^2 \theta_{12}$ and 
the atmospheric mixing angle in $\sin^2 \theta_{23}$; the larger the former,
the smaller the latter.  The upper bands in Figs. 4(d) and (e) are obtained 
when both parameters $a$ and $b$ have positive imaginary values and yield a 
positive Dirac phase $\delta_{CP}$, while the lower bands arise from negative 
imaginary values for $a$ and $b$.  Note that for the restriction 
$\eta_B \geq 1.0 \times 10^{-10}$, the leptonic CP phase lies in the two 
ranges, $\pm(50^\circ, 80^\circ)$. 

The distribution of points in $\eta_B$ and $\sin^2 \theta_{23}$ which survive
the cut of $\eta_B \geq 1.0 \times 10^{-10}$ is shown in Fig. 4(f).  Several
points with $\eta_B \simeq 6.0 \times 10^{-10}$ are observed to be clustered 
around $\sin^2 \theta_{23} \simeq 0.6$.  For these few points, values of 
$c \sim \pm a$ seem to be preferred.  This suggested that we construct 
scatter plots with all three parameters, $a,\ b$ and $c$ taken to be imaginary
with the restriction $c = a$.  The corresponding plots are shown in Fig. 5,
where 100,000 sets of points are thrown and 381 points survive.  It is rather 
remarkable that the atmospheric neutrino mixing plane in Fig. 5(a) is 
restricted to a short arc centered at $\Delta m^2_{32} = 2.45 \times 10^{-3}\ 
{\rm eV^2}$ and with $\sin^2 2\theta_{\rm atm}$ lying in the range 
(0.985, 0.995), as $0.45 \leq \sin^2 \theta_{23} \leq 0.55$ with $2.30 \times 
10^{-3} \leq \Delta m^2_{32} \leq 2.65 \times 10^{-3}\ {\rm eV^2}$ for our 
choice of $\Lambda_R$.  The solar mixing plane in Fig. 5(b) is also somewhat 
truncated at the lower end of the spectrum as now $\sin^2 2\theta_{\rm sol} 
\gsim 0.80$.  The bands in Figs. 5(c)-(e) are similar to those in Fig. 4(c)-(e)
but again somewhat truncated by the maximum value of $\theta_{23}$.  The 
relevant values of $\sin^2 \theta_{13}$ now lie in the interval 
(0.0020, 0.0031).  The CP phase is bounded by $\pm(60^\circ, 86^\circ)$
in Fig. 5(f).  

With the strong correlation of the mixing angles noted above, $\eta_B$ is now
directly related to the mixing angle $\theta_{23}$ in Fig. 5(f), as the random 
scattering in Fig. 4(f) has disappeared.  The observed baryon 
asymmetry value of $6.2 \times 10^{-10}$ is obtained for $\sin^2 \theta_{23} 
\simeq 0.55$, which implies $\tan^2 \theta_{12} \simeq 0.38$, 
$\sin^2 \theta_{13} \simeq 0.0021$ and $\delta_{CP} \simeq \pm 60^\circ$.  For 
this case the sum of $\theta_{12}$ and the Cabibbo angle is given by 
$\theta_C + \theta_{12} = 44.8^\circ$, very close to the value of $45^\circ$
which is referred to as quark-lepton complementarity in the literature 
\cite{compl}.  On the other hand, maximal atmospheric neutrino mixing with 
$\sin^2 \theta_{23} = 0.5$ corresponds to 
$\tan^2 \theta_{12} = 0.44$, $\sin^2 \theta_{13} \simeq 0.0024$, $\delta_{CP} 
= \pm 75^\circ$, and $\eta_B = 3.8 \times 10^{-10}$.

As an example of a complete solution, we present the results for a special 
case which arises for the choice of right-handed Majorana neutrino parameters: 
$a = 0.5828i,\ b = 1.7670i,\ c = 0.5828i$ with $\Lambda_R = 2.85 
\times 10^{14}$ GeV, $\eta = 1.1 \times 10^{-5},\ \delta_N = - 1.0 \times 
10^{-5}$, and $\delta'_N = - 1.5 \times 10^{-5}$.  One obtains 
$\eta_B = 6.2 \times 10^{-10}$ with the following mixing parameters
\begin{equation}
\begin{array}{rlrl}
  \Delta m^2_{32} &= 2.30 \times 10^{-3}\ {\rm eV}^2,\qquad & 
       \sin^2 \theta_{23} &= 0.550,\\
  \Delta m^2_{21} &= 7.71 \times 10^{-5}\ {\rm eV}^2,\qquad & 
       \tan^2 \theta_{12} &= 0.388,\\
  \sin^2 \theta_{13} &= 0.0022,\qquad &
       \delta_{CP} &= 63^\circ,\\
  \chi_1 &= -55.3^\circ,\qquad & \chi_2 &= 31.8^\circ,\\
\label{speccase}
\end{array}
\end{equation}
where $\chi_1$ and $\chi_2$ are the Majorana phases in the Majorana phase
matrix, $\Phi = {\rm diag}(\exp i\chi_1,\ \exp i\chi_2,\ 1)$.  For this case, 
$m_3 = 48.9$ meV, $m_2 = 9.30$ meV, $m_1 = 3.05$ meV, $M_3 = 2.91 \times 
10^{14}$ GeV, $M_1 \sim M_2 \sim 5.40 \times 10^8$ GeV, $M_2 - M_1 
= 1.06 \times 10^3$ GeV, and $\Gamma_1 = 2.07 \times 10^3$ GeV where 
$\Gamma_1$ is the width of the lightest right-handed state.

Further study reveals that the narrow band structures in Fig. 5, obtained
with $a,\ b$ and $c$ pure imaginary and $c = a$, arise because $b$ is limited
to the very narrow ranges, $b = \pm (1.7\ {\rm to}\ 2.1)i$, for all 381 points 
which satisfy the mixing parameter constraints imposed in 
Eq. (\ref{eq:nudata}).  On the other hand, the ratio $a/b$ varies in the range
$a/b = (0.33\ {\rm to}\ 0.53)$ along each band as $\sin^2 \theta_{23}$ 
decreases from 0.55 to 0.45.  The pair of slightly displaced bands in Fig. 5(d)
is associated with the choices of positive or negative imaginary values for 
$a$ and $b$.  Clearly the imposition of these restrictions 
for the $a,\ b,\ c$ parameter set has greatly enhanced the probability that
the baryon asymmetry will lie in or near to its observed value.  This makes
for a rather compelling scenario within the model.
 
The question arises as to how likely is it that such restricted values 
should apply for the Majorana mass parameters.  It is easy to see from the 
Froggatt-Nielsen diagrams for the Majorana mass matrix that if only one 
Higgs VEV is responsible for breaking lepton number, by the $U(1) \times 
Z_2 \times Z_2$ family symmetry of the model only one diagram will 
contribute to each matrix element, the structure of $M_R$ will be purely 
geometrical and its determinant will vanish, cf. Ref. \cite{abLMA}.  With 
$b \neq c = a$, only one diagram contributes to each of the 11, 13, and 31 
matrix elements, while two or more diagrams will be required for the 
12 and 21 elements.  This is the simplest possibility for a nonvanishing
determinant, as required for the seesaw mechanism.  The fact that all three
parameters are purely imaginary makes for a maximum CP violation in the 
right-handed Majorana mass matrix as all other parameters are real.

To understand the apparent difficulty in reaching values in the observed 
range of $\eta_B \simeq 6.2 \times 10^{-10}$, we note the following.  
The scenario of resonant leptogenesis \cite{reslepto} arises naturally in 
the model, for the two heavy right-handed Majorana 
neutrinos have nearly equal masses and nearly opposite CP-parity.  However, 
one must impose the condition that the mass difference of these two heavy 
Majorana states be at least as large as the half-width of either state 
for the resonance formula to make sense.  This forced us to raise the 
magnitudes of $\eta,\ \delta_N$ and $\delta'_N$ in the Dirac neutrino mass 
matrix.  For smaller values of these parameters than those chosen in the 
relevant figures, fewer points reach the $10^{-10}$ range of $\eta_B$, as the  
range of $\sin^2 \theta_{23}$ is shifted lower with more points occuring 
below 0.50 than above 0.50.  It is interesting that in going from the  
set of matrices with the original choice of $\eta,\ \delta_N$ and $\delta'_N$ 
in Fig. 1 to the new choice of values in Figs. 2 - 5, the $V_{us}$ mixing 
parameter in the quark mixing matrix is raised from 0.2220 to 0.2240, in better
keeping with the new observed value which is preferred for unitarity reasons.  
The one negative effect in making this change is that the up quark mass is 
raised from 3.5 MeV to 5.5 MeV after evolution from the GUT scale, which is 
above the preferred range.

While the condition for thermal resonant leptogenesis is easily
satisfied for such right-handed Majorana masses as in the example given
above, the overproduction of gravitinos in the early universe is a 
problem in the supergravity scenario of SUSY breaking; this can be 
alleviated if the SUSY breaking occurs via the gauge-mediated scenario 
\cite{gravitino}.  As illustrated in Ref. \cite{lepto}, this problem can
be completely avoided and satisfactory leptogenesis and baryon asymmetry 
obtained, if the model is expanded to include three additional intermediate 
mass scale singlets which lead to a double seesaw mechanism.  However, with 
this expanded seesaw mechanism many more parameters are introduced, and the
model becomes much less predictive.

\section{CONCLUDING REMARKS}
  
From the above results we can conclude that the model with a normal mass 
hierarchy can easily fit the presently observed neutrino mixing data.  
As originally proposed, eight of the twelve model parameters have been fixed 
by the quark and lepton masses and quark mixing data, while the remaining 
four (three complex and one real) parameters of the right-handed Majorana 
mass matrix are varied.  The real $\Lambda_R$ parameter centers the 
$\Delta m^2_{32}$ model spectrum on the observed $90\%$ confidence level 
interval given in Eq. (\ref{eq:nudata}).  For the best fit point of 
$\Delta m^2_{32} = 2.4 \times 10^{-3}\ {\rm eV}^2$, the predicted atmospheric 
neutrino mixing angle is nearly maximal, corresponding to 
$\sin^2 2\theta_{23} \gsim~0.98$ in agreement with the observed best fit 
value of 1.00.  No such preference for a particular value of 
$\tan^2 \theta_{12}$ in the allowed range is found.  The predicted values 
for $\sin^2 \theta_{13}$ are of special interest.  There we found a range of 
$1 \times 10^{-5} \lsim \sin^2 \theta_{13} \lsim 0.01$ which lies noticeably 
lower than that predicted by most other models where values very close to the 
upper bound of 0.04 are more generally expected \cite{13models}.  This is true 
not only for the nearly conserved $L_e - L_\mu - L_\tau$ models, but also for 
the $SO(10)$ models with symmetric ${\bf 126}_H$ and $\overline{\bf 126}_H$ 
Higgs representations.  The lower values predicted in our model have relevance 
for the future study of neutrino oscillations with reactor beams.  If 
$\bar{\nu}_e$ disappearance is observed very close to the present CHOOZ bound 
by the next generation reactor experiments, the $SO(10)$ model under 
consideration will be ruled out.  On the other hand, it is interesting to note 
that if future reactor experiments do not see some $\bar{\nu}_e$ disappearance,
the model predicts it will certainly be seen with superbeam and/or neutrino 
factory experiments.  While there remains much interest in determining the 
Dirac phase $\delta_{CP}$ for $CP$ violation, it is rather surprising that 
the model is able to accommodate almost any value.  

With the original Dirac neutrino mass parameters, the resonance leptogenesis
inherent in the model generated a baryon asymmetry which falls more than two
orders of magnitude short of the observed value.  In an effort to increase
the effect, the original very small $\eta$ parameter was adjusted and two 
new equally small parameters, $\delta_N$ and $\delta'_N$, were added to the 
Dirac neutrino mass matrix.  This helped to recover two missing orders of 
magnitude for $\eta_B$, provided the three complex parameters in the 
right-handed Majorana mass matrix were chosen very near to or on their 
imaginary axes.  By selecting only those points which satisfied the observed 
neutrino mixing results and for which $\eta_B \gsim 1.0 \times 10^{-10}$, 
we observed that the mixing angles lie along very narrow bands.  

These bands were further dramatically tightened by choosing not only $a,\ b$ 
and $c$ purely imaginary, but restricting $c$ to be equal to $a$.  For this 
situation a limited range of $0.45 \leq \sin^2 \theta_{23} \leq 0.55$ was 
found over which $\tan^2 \theta_{12}$ falls from 0.50 to 0.38, $\sin^2 
\theta_{13}$ varies from 0.0031 to 0.0020, and the Dirac CP phase varies from 
$\pm(85^\circ \ {\rm to}\ 60^\circ)$.  On the other hand, a narrow band also
developed for the baryon asymmetry due to the strong mixing angle and 
CP phase correlations, as it rises from $(2.7\ {\rm to}\ 6.3) \times 10^{-10}$
with increasing $\sin^2 \theta_{23}$.  The appearance of such narrow bands 
is directly related to the very narrow acceptable range for $b$, while the 
ratio $a/b$ varies along the length of each band.  We argued that the $c = a$ 
restriction can be understood in terms of one Froggatt-Nielsen diagram for 
each of the 11, 13, and 31 matrix elements of $M_R$, while two such diagrams 
can contribute to the 12 and 21 elements.  All these results are of 
considerable interest and will severely test the model in the future. 

Since the light neutrino mass spectrum has a normal hierarchy, the model 
predicts no neutrinoless double beta decay should be observed in the near
future.  In fact, for all choices of the model parameters which lead to 
acceptable ranges for the mixing parameters of Eq. (\ref{eq:nudata}), we 
find the effective mass is $m_{\beta\beta} \sim 0.4 \times 10^{-3}\ {\rm eV}$
\cite{typeI}, three orders of magnitude below the present bound 
\cite{mbetabeta}.

In conclusion, we address briefly the issue of lepton flavor violation in the 
decay process $\mu \rightarrow e + \gamma$.  Jankowski and Maybury \cite{jm} 
have studied this particular class of $SO(10)$ models with lopsided charged 
lepton mass matrices in order to determine the expected branching ratio for 
the above decay, under the assumption that the supersymmetric grand unified 
model breaks directly to the constrained minimal supersymmetric standard model.
Given the combined constraints on the CMSSM parameters from direct searches 
and from the WMAP satellite observations \cite{etaB,WMAP}, they find that for 
a low value of $\tan \beta \simeq 5$ preferred by the model, the branching 
ratio should be of $\mathcal{O}$($10^{-12}$) and very close to the present 
upper limit of $1.2 \times 10^{-11}$ \cite{muegamma}.  Hence the $\mu 
\rightarrow e + \gamma$ decay should be observed in the new MEG experiment 
\cite{MEG} which is expected to start taking data in 2006.  This prediction is 
based on the slepton masses being in the 125 - 150 GeV range, while the 
neutralino masses are around 300 - 500 GeV.  These ranges are also favored in 
CMSSM models by the recent study of the best chi-squared fits to all the data 
with $\tan \beta \simeq 10$ by Ellis et al. \cite{CMSSM}.  On the other hand, 
minimal $SO(10)$ models leading to symmetric or antisymmetric mass matrix 
elements favor a smaller lepton-violating branching ratio of 
$\mathcal{O}$($10^{-13}$) or less \cite{flavor}.  The coming MEG 
experiment which is expected to be sensitive down to the $10^{-14}$ branching 
ratio level may thus be able to rule out one of the two classes of minimal 
$SO(10)$ models before the new reactor experiments can be launched.\\   

The author thanks Uli Haisch and Enrico Lunghi for their help with
the scatter plots.  The author also thanks the Theory Group at Fermilab for 
its kind hospitality.  Fermilab is operated by Universities Research 
Association Inc. under contract No. DE-AC02-76CH03000 with the Department 
of Energy. 
%

%
%
\newpage
\begin{figure*}[t]
\vspace*{-1in}
\hspace*{-1in} \includegraphics*{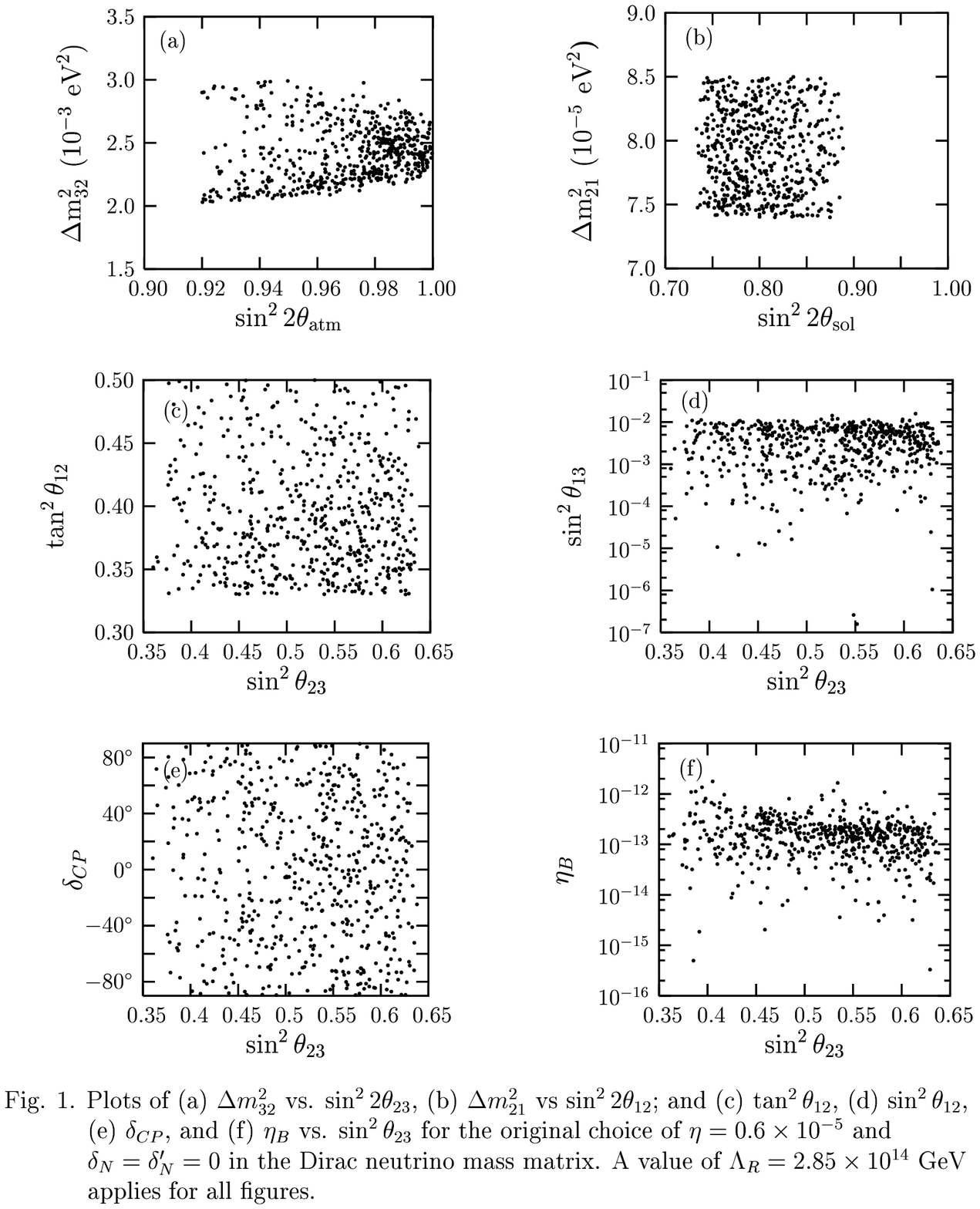}
\end{figure*}
\newpage
\begin{figure*}[t]
\vspace*{-1in}
\hspace*{-1in} \includegraphics*{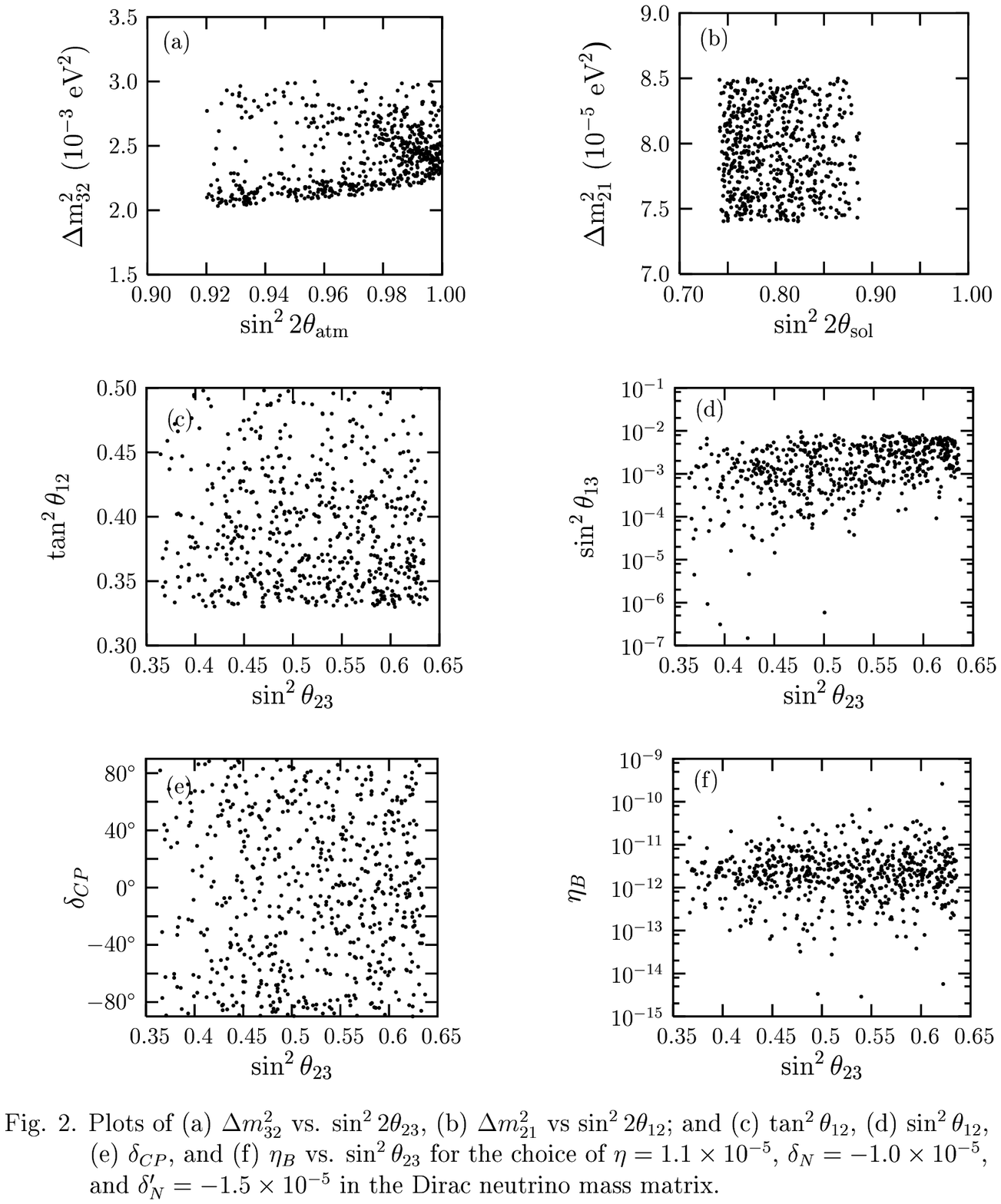}
\end{figure*}
\newpage
\begin{figure*}[t]
\vspace*{-1in}
\hspace*{-1in} \includegraphics*{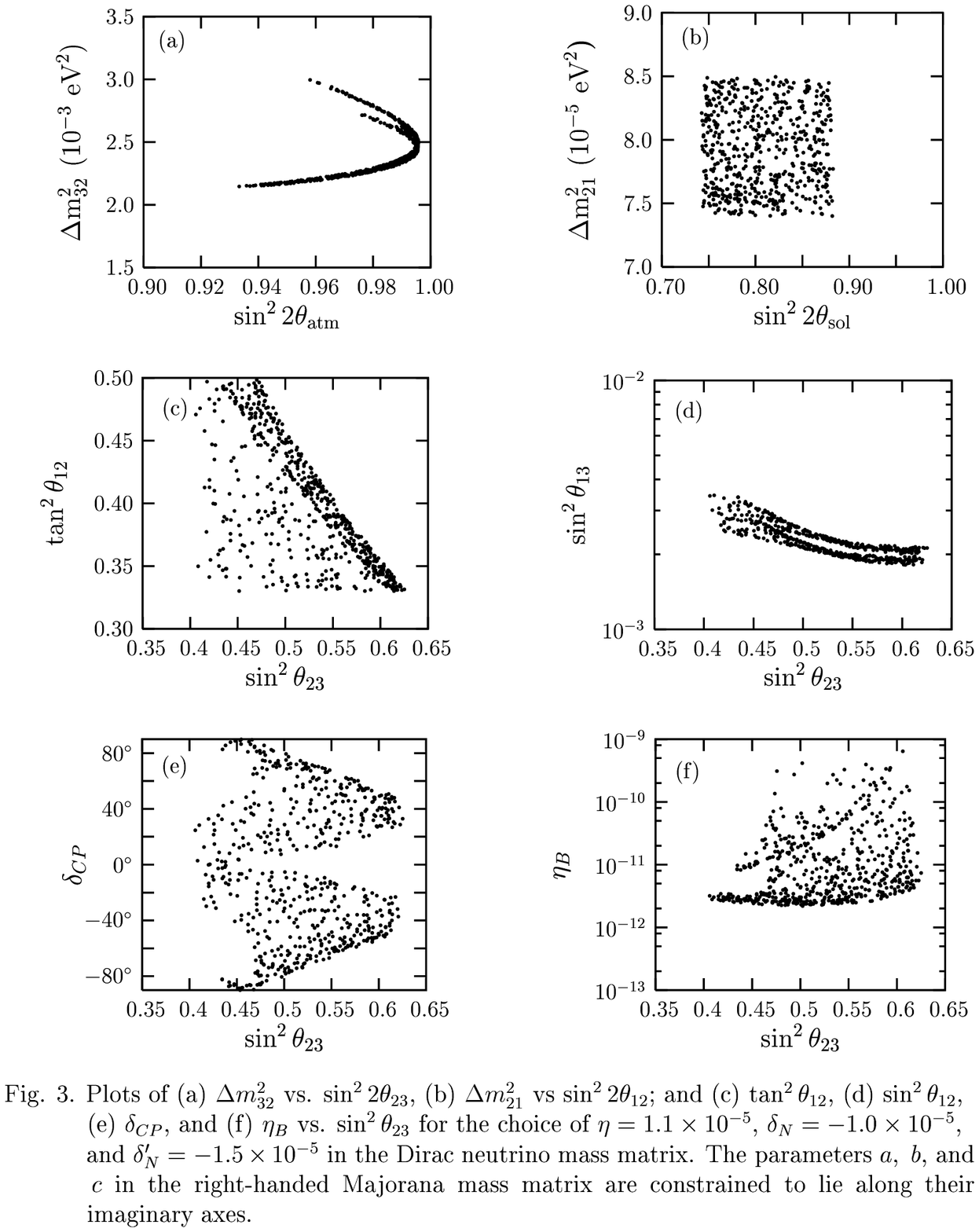}
\end{figure*}
\newpage
\begin{figure*}[t]
\vspace*{-1in}
\hspace*{-1in} \includegraphics*{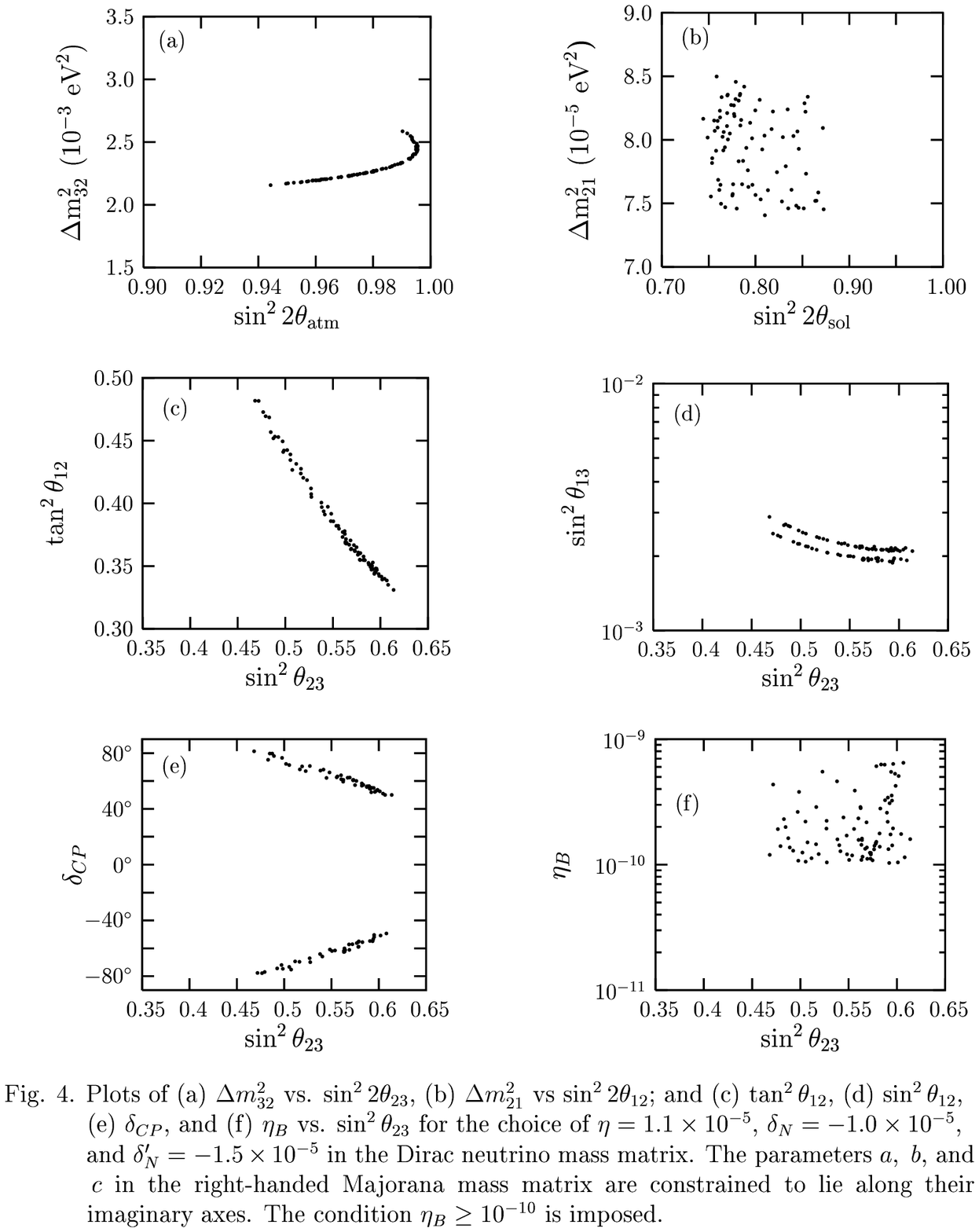}
\end{figure*}
\newpage
\begin{figure*}[t]
\vspace*{-1in}
\hspace*{-1in} \includegraphics*{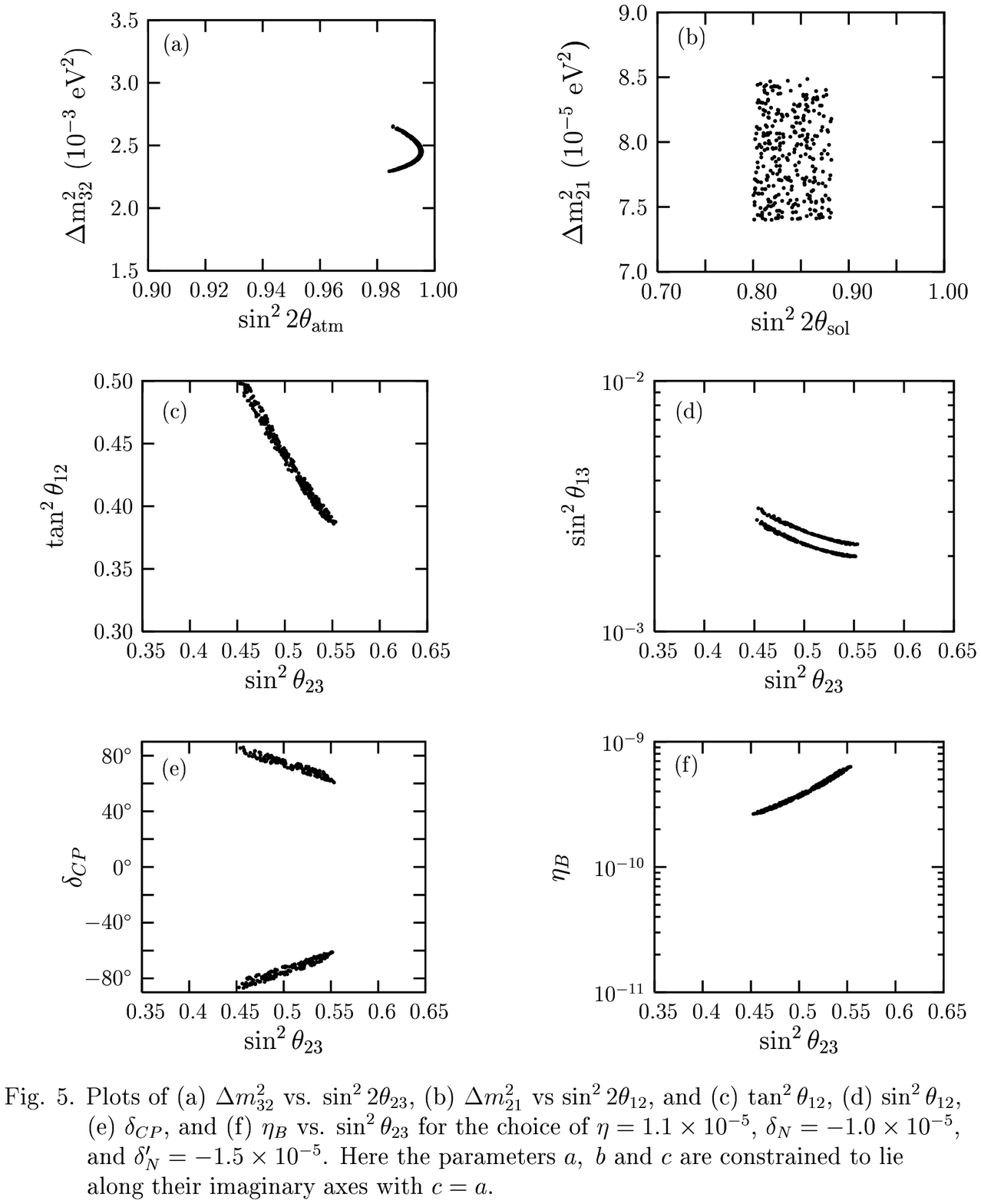}
\end{figure*}

\end{document}